\documentclass[twoside,aps]{revtex4}

\usepackage{graphicx} 
\usepackage{color} 
\usepackage{rotating}
\usepackage{epsfig}

\newcommand{\ket}[1]{\left|#1\right\rangle}

\begin{document}

\title{Surface code quantum error correction incorporating accurate error propagation}

\author{Austin G. Fowler, David S. Wang, Lloyd C. L. Hollenberg}
\affiliation{Centre for Quantum Computer Technology, University of
Melbourne, Victoria, Australia}

\date{\today}

\begin{abstract}
The surface code is a powerful quantum error correcting code that
can be defined on a 2-D square lattice of qubits with only nearest
neighbor interactions. Syndrome and data qubits form a checkerboard
pattern. Information about errors is obtained by repeatedly
measuring each syndrome qubit after appropriate interaction with its
four nearest neighbor data qubits. Changes in the measurement value
indicate the presence of chains of errors in space and time. The
standard method of determining operations likely to return the code
to its error-free state is to use the minimum weight matching
algorithm to connect pairs of measurement changes with chains of
corrections such that the minimum total number of corrections is
used. Prior work has not taken into account the propagation of
errors in space and time by the two-qubit interactions. We show that
taking this into account leads to a quadratic improvement of the
logical error rate.
\end{abstract}

\maketitle

\section{Introduction}

The idea of manipulating quantum systems to perform computation was
first proposed by Feynman in 1982 \cite{Feyn82}. Serious interest in
making this vision a reality followed the invention of Shor's
factoring algorithm in 1994 \cite{Shor94b}. Initial concerns that
unfeasible physical control was required were tempered by the
invention of quantum error correction (QEC) in 1995 \cite{Shor95}
and proof of the threshold theorem in 1996 \cite{Knil96b, Ahar97}.
The threshold theorem showed that arbitrarily large quantum
computations can be performed arbitrarily reliably provided the
error rates of the various components in the quantum computer are
all below some fixed threshold error rate $p_{\rm th}$.

Early work \cite{Ahar99} put the value of $p_{\rm th}$ at
approximately $10^{-6}$. The ability to interact pairs of qubits
separated by arbitrarily large distances was also demanded. The
current record highest $p_{\rm th}$ of approximately 3\% is shared
by two quantum computing schemes \cite{Knil04c, Fuji09}. Both of
these schemes, however, still require the ability to interact pairs
of qubits separated by arbitrarily large distances. This is a major
barrier to implementation. Studies of the performance of such error
correction techniques when they are restricted to 1-D, quasi-1-D and
2-D lattices of qubits with only nearest neighbor interactions
\cite{Step09, Step07a, Fowl07, Helm09, Szko04, Svor04} have found
that the threshold error rate drops to the $10^{-5}$ level or below.

Topological approaches to quantum error correction \cite{Brav98,
Denn02, Raus06, Raus07, Raus07d, Bomb06b, Bomb06, Bomb07b, Bomb07c,
Bomb07, Bomb08, Katz10, Stac09, Stac09b, Ducl09, Fowl08, Fowl09,
Fowl09d, Wang09, Wang09b, Devi09} can be implemented optimally using
only nearest neighbor interactions. Kitaev's surface code
\cite{Brav98}, which makes use of a 2-D square lattice of qubits
with nearest neighbor interactions, possesses the highest threshold
error rate of the known topological schemes at approximately 0.75\%
\cite{Raus07, Fowl08, Wang09}. This is not far below the record
threshold error rate and above some existing experimental error
rates \cite{DiCa09, Pass09}. In recent years, a number of quantum
computer designs based on this form of error correction have been
devised \cite{Stoc08, Devi07, Step08, Devi08, VanM09, DiVi09} and an
experimental demonstration using linear optics has been performed
\cite{Gao09}. Many technologies are being developed that are in
principle capable of implementing a 2-D lattice of qubits with
nearest neighbor interactions, including ion traps \cite{Blak09,
Leib09, Home09, Hann09, Amin09}, neutral atom chips \cite{Fort07,
Ried10}, optical lattices \cite{Bloc08, Beck09}, superconductors
\cite{DiCa09, Bial09} quantum dots \cite{Tayl05}, donors in silicon
\cite{Holl06} and electrons on liquid helium \cite{Lyon06}.

A number of open problems remain concerning the best way to perform
the classical processing associated with surface code QEC. In this
paper, we address a major failing of prior approaches \cite{Fowl08,
Wang09, Ducl09}, which did not take into account the propagation of
errors as a result of the two-qubit interactions required to detect
errors. We show that, for a given lattice size, carefully taking
into account error propagation results in a quadratic improvement of
the logical error rate.

The discussion is organized as follows. In Section~\ref{The surface
code}, the surface code is briefly reviewed. Section~\ref{Standard
surface code QEC} discusses the standard surface code QEC procedure,
its limitations, and numerical results explicitly demonstrating
these limitations. Section~\ref{Improved surface code QEC} describes
our modification to the standard procedure, its advantages, and
numerical results explicitly demonstrating these advantages.
Section~\ref{Conclusion} concludes with a summary of our results.

\section{The surface code}
\label{The surface code}

In this Section, we briefly describe the surface code, focusing
primarily on its stabilizers and the quantum circuits required to
measure them. A full description of surface code quantum computing
can be found in \cite{Fowl08}.

\begin{figure}
\begin{center}
\resizebox{140mm}{!}{\includegraphics{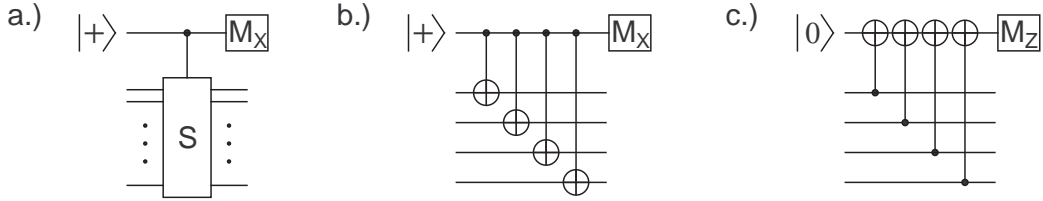}}
\end{center}
\caption{a.) General circuit determining the sign of a stabilizer
$S$. b.) Circuit determining the sign of a stabilizer $XXXX$. c.)
Circuit determining the sign of a stabilizer
$ZZZZ$.}\label{fig:figure1}
\end{figure}

\begin{figure}
\begin{center}
\resizebox{180mm}{!}{\includegraphics{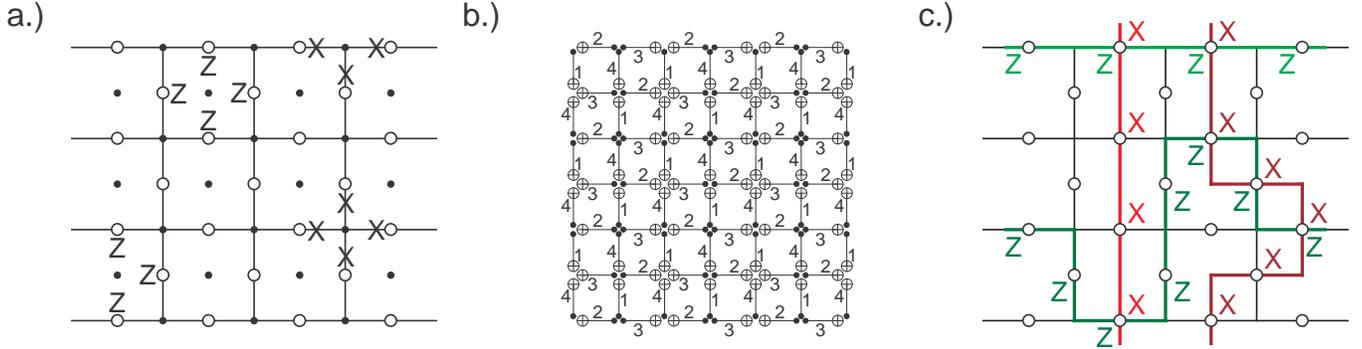}}
\end{center}
\caption{a.) 2-D lattice of data (open circles) and syndrome (dots)
qubits along with examples of the data qubit stabilizers. b.)
Sequence of CNOTs permitting simultaneous measurement of all
stabilizers. The numbers indicate the relative gate order. c.)
Examples of logical operators.}\label{fig:figure2c}
\end{figure}

A stabilizer of a state $\ket{\Psi}$ is an operator $S$ such that
$S\ket{\Psi} = \ket{\Psi}$. Any error $E$ that anticommutes with $S$
can be readily detected since $SE\ket{\Psi} = -ES\ket{\Psi} =
-E\ket{\Psi}$. A generic circuit capable of determining the sign of
a stabilizer is shown in Fig.~\ref{fig:figure1}a. Surface code
stabilizers have the form $XXXX$, $XXX$, $ZZZZ$ or $ZZZ$, as shown
for the specific case in Fig.~\ref{fig:figure2c}a. Circuits capable
of measuring such stabilizers are shown in
Figs.~\ref{fig:figure1}a-b. An appropriate sequence of two-qubit
gates to use when measuring all stabilizers across the lattice
simultaneously is shown in Fig.~\ref{fig:figure2c}b.

The lattice of Fig.~\ref{fig:figure2c}a is capable of protecting a
single qubit of information from errors. This protected qubit is
called a logical qubit and can be read out in the logical $X/Z$
basis by measuring all data qubits in the physical $X/Z$ basis.
After error correction, which we shall describe in the next Section,
the logical measurement result is equal to the product of $X/Z$
measurement results along paths connecting boundaries as shown in
Fig.~\ref{fig:figure2c}c. The distance $d$ of the code is equal to
the shortest of these paths. Fig.~\ref{fig:figure2c}a is an example
of a distance $d=4$ code.

\section{Standard surface code QEC}
\label{Standard surface code QEC}

\begin{figure}
\begin{center}
\resizebox{180mm}{!}{\includegraphics{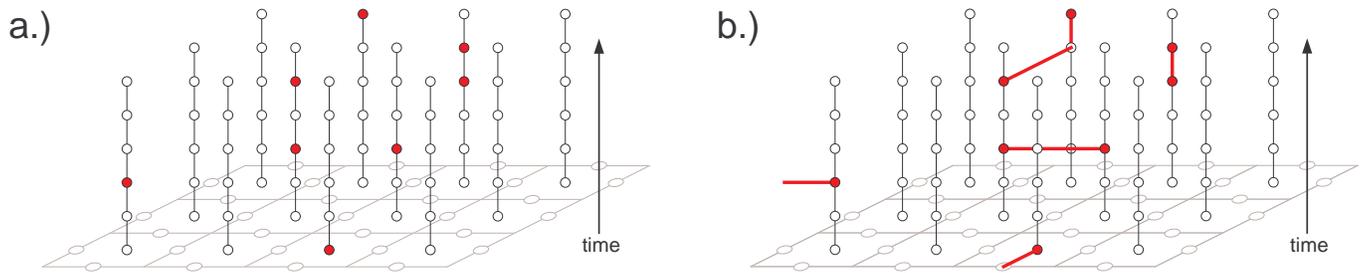}}
\end{center}
\caption{a.) Locations in space and time, indicated by red dots,
where and when the reported syndrome is different from that in the
previous time step.  Note that this is not a three-dimensional
physical structure, just a three-dimensional classical data
structure.  b.) Optimal matching highly likely to lead to a
significant reduction of the number of errors if bit-flips are
applied to the spacelike edges.}\label{fig:matching_example}
\end{figure}

Repeatedly executing the gate sequence shown in
Fig.~\ref{fig:figure2c}b, along with syndrome qubit initialization
and measurement, generates data points in space and time where the
measurement values change. These are called syndrome changes.
Fig.~\ref{fig:matching_example}a shows an example. The standard
method of surface code QEC effectively only considers errors
occurring on data qubits at the same time as the syndrome qubits are
being measured and errors on the syndrome qubits just before they
measured. Such errors are not propagated by the two-qubit gates. A
single data qubit error results in two syndrome changes adjacent in
space. A single syndrome qubit error results in two syndrome changes
adjacent in time. The separation $s$ of two syndrome changes
occurring at space-time locations $(i_1, j_1, t_1)$, $(i_2, j_2,
t_2)$ is defined to be $s=|i_1-i_2|+|j_1-j_2|+|t_1-t_2|$. In other
words, the standard approach assumes that an error chain containing
at least $s$ errors must occur to produce the observed syndrome
changes. The minimum weight matching algorithm \cite{Edmo65a,
Edmo65b} is used to process the list of space-time locations,
matching pairs of locations such that the total of all separations
is a minimum. An example is shown in
Fig.~\ref{fig:matching_example}b.

\begin{figure}
\begin{center}
\resizebox{180mm}{!}{\includegraphics{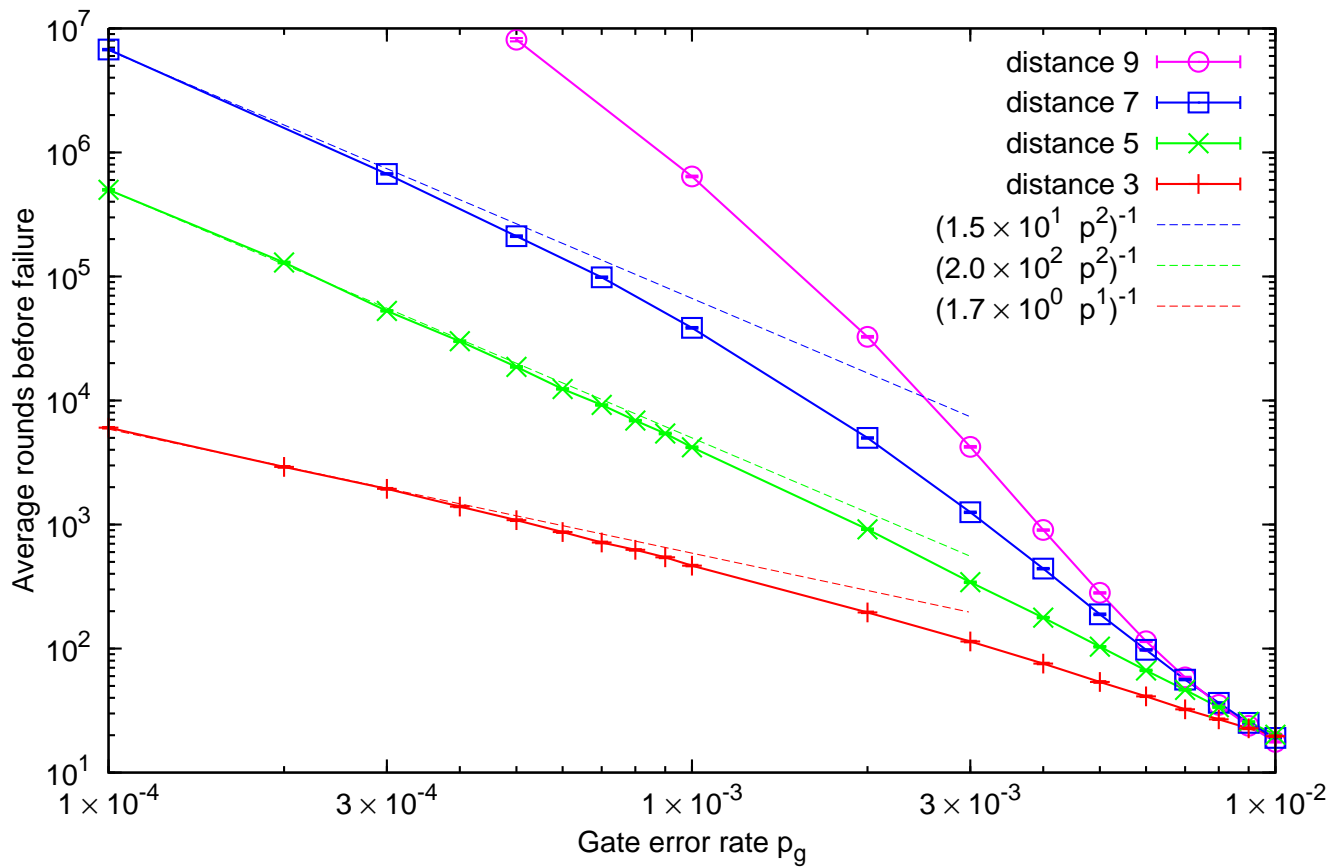}}
\end{center}
\caption{Average number of rounds of error correction before the
logical qubit is corrupted as a function of the physical gate error
rate $p_g$ and code distance $d$ when the standard measure of
separation $s$ of syndrome changes is used. Asymptotic curves have
been included.}\label{fig:dt_simple}
\end{figure}

The performance of this approach is shown in
Fig.~\ref{fig:dt_simple}. The error rates of initialization,
measurement, memory and two-qubit gates are all set to equal $p_g$.
The number of rounds of syndrome measurement that can be performed,
on average, before a logical error occurs is plotted for a range of
lattice sizes $d$ and gate error rates $p_g$. It can be seen that
something is seriously wrong. At low physical error rates, a
distance $d=3$ lattice has a logical error rate proportional to
$p_g$, meaning it offers no genuine error correction ability. Both
$d=5$ and $d=7$ are only capable of reliably correcting a single
error.

\begin{figure}
\begin{center}
\resizebox{100mm}{!}{\includegraphics{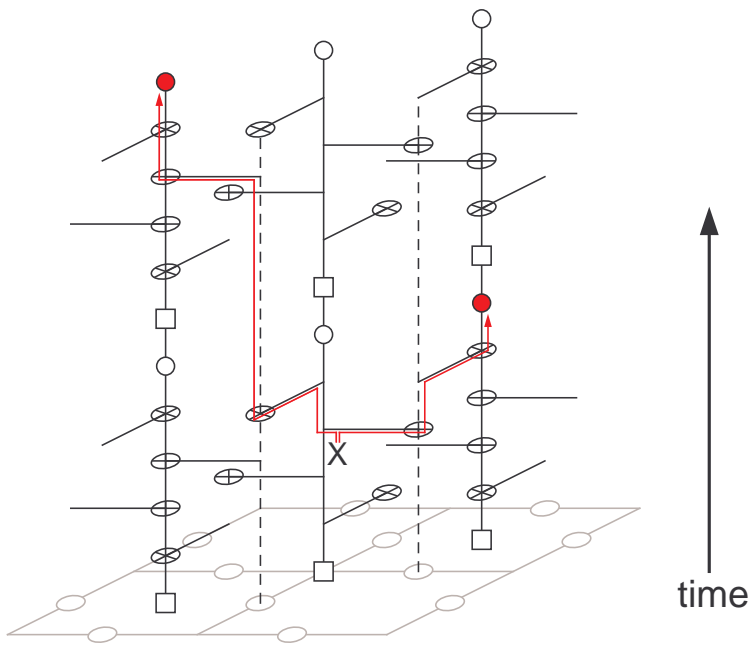}}
\end{center}
\caption{Errors on syndrome qubits can propagate to two locations
separated by two units of space and one unit of
time.}\label{fig:syndrome_error}
\end{figure}

\begin{figure}
\begin{center}
\resizebox{70mm}{!}{\includegraphics{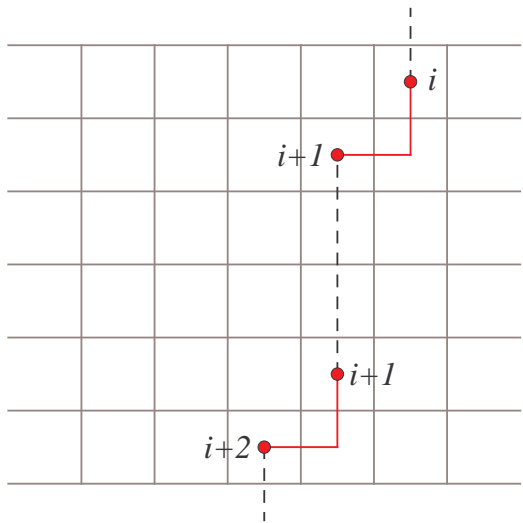}}
\end{center}
\caption{Two errors of the form shown in
Fig.~\ref{fig:syndrome_error} can cause a distance $d=7$ code to
fail when simple syndrome change separations are used. The dashed
lines indicate five operators that will be suggested as appropriate
corrections but will in fact form a logical
error.}\label{fig:failure_example2}
\end{figure}

The reason the error correction performs suboptimally at low error
rates is illustrated in Fig.~\ref{fig:syndrome_error}. A single
error occurring halfway through syndrome measurement can cause a
pair of syndrome changes separated by two units of space and one of
time. A chain of such errors as shown in
Fig.~\ref{fig:failure_example2} will be matched as shown in dashed
lines, resulting in a logical error. A $d\times d$ lattice can thus
only cope with $\lfloor (d-1)/4 \rfloor$ errors in the worst case
using this error correction method.

\section{Improved surface code QEC}
\label{Improved surface code QEC}

A $d\times d$ lattice should be able to cope with $\lfloor (d-1)/2
\rfloor$ errors in the worst case. We achieve this by modifying the
way in which the separation is calculated between a pair of syndrome
changes.

Let us consider a slightly more general syndrome measurement
procedure than that shown in Fig.~\ref{fig:figure2c}b. Let us assume
that some of the two-qubit interactions are significantly slower
than others. Some syndrome measurements may be performed more
frequently than others. We permit the order of interaction to vary
dynamically to enable syndrome measurements to be performed as
frequently as possible. We imposed the necessary condition that it
must be possible to say whether the gates associated with a given
syndrome measurement occurred strictly before or after those
associated with any other syndrome measurement.

\begin{figure}
\begin{center}
\resizebox{160mm}{!}{\includegraphics{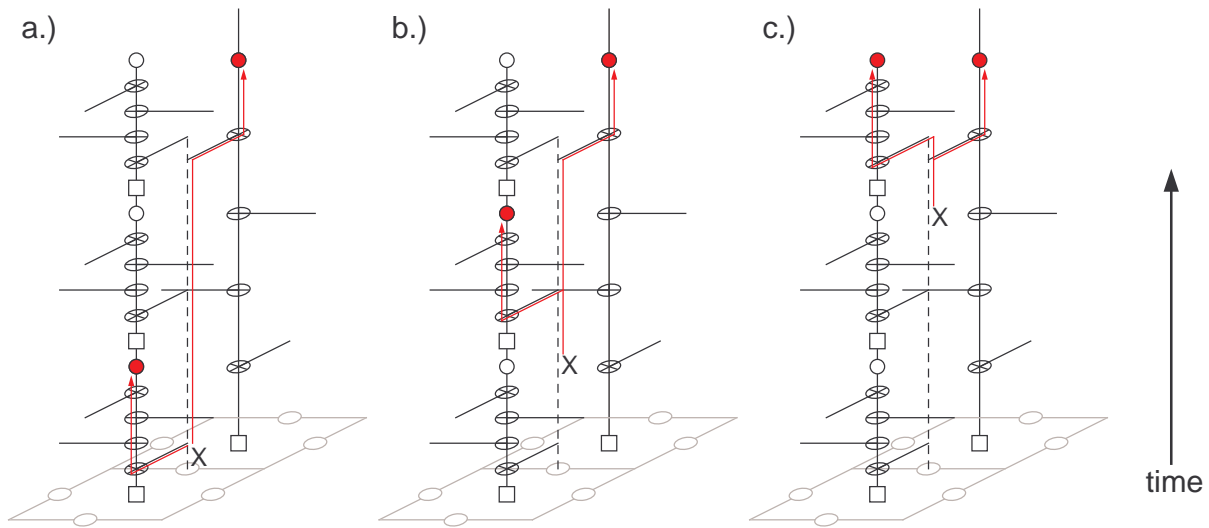}}
\end{center}
\caption{Example of propagation of data errors when one syndrome is
measured more frequently than another. Changed syndromes are
indicated in red.}\label{fig:data_error}
\end{figure}

Consider a pair of neighboring $Z$ stabilizers sharing a single data
qubit. Let us imagine that one stabilizer is measured more
frequently than the other. $X$ errors occurring on the data qubit at
different times will be detected by the adjacent stabilizer
measurements in a number of different ways as shown in
Fig.~\ref{fig:data_error}. These are all single-error processes. As
such, the pairs of changed syndromes indicated in red should all be
connected. These connections, along with vertical connections
between same site syndrome measurements and connections of the form
shown in Fig.~\ref{fig:syndrome_error} will be used to determine the
correct separation of a given pair of syndrome changes. The
separation of a given pair of syndrome changes is defined to be the
minimum number of connections in any path connecting them.

It may seem that there must be additional connections to cover the
case of two-qubit interactions suffering a two-qubit error, however
this is not the case. Two-qubit $XX$ and $ZZ$ errors are equivalent
to single $X$ or $Z$ errors occurring before the two-qubit gate.
Two-qubit errors consisting of both $X$ and $Z$ operators give rise
to propagated errors that are handled independently by the two types
of error correction.

\begin{figure}
\begin{center}
\resizebox{180mm}{!}{\epsfig{figure=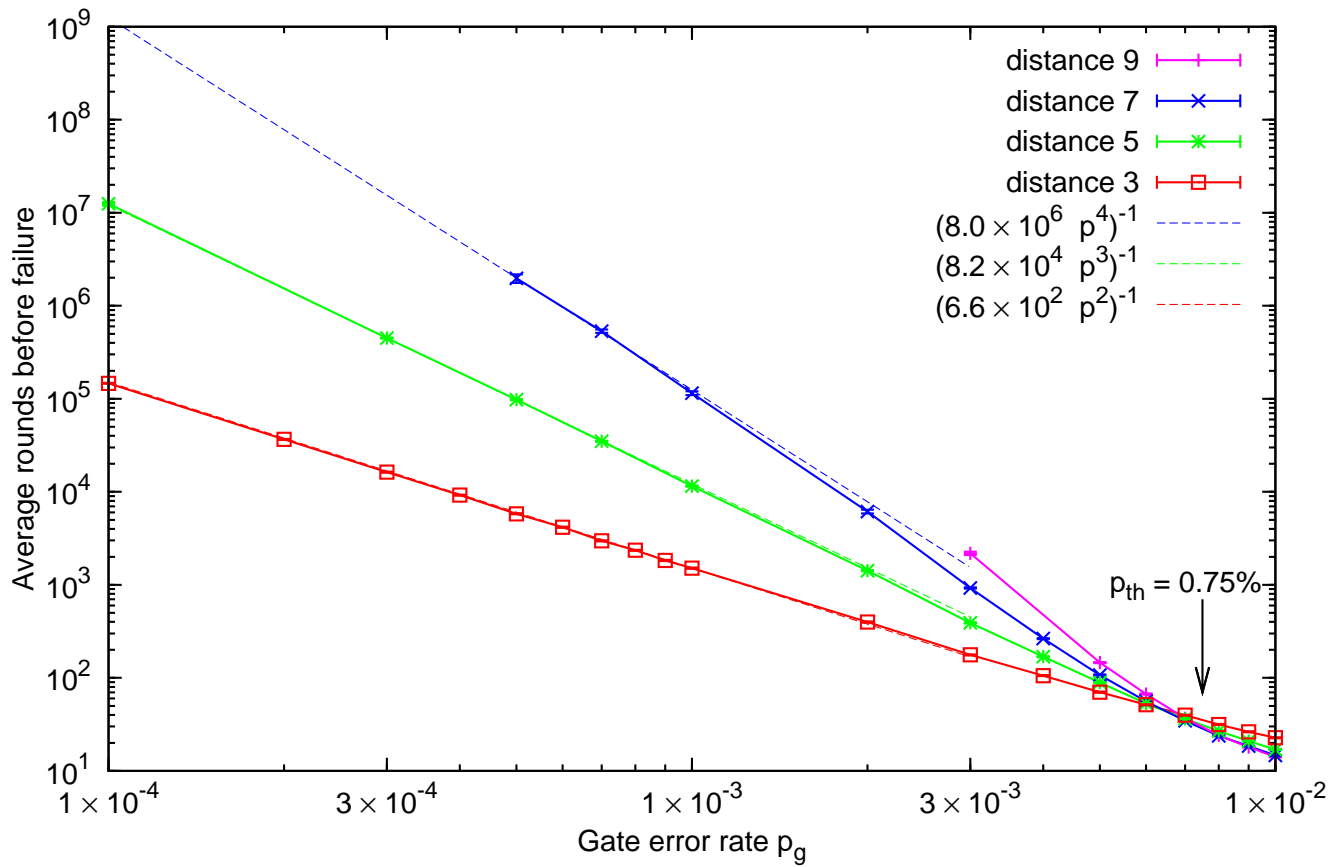}}
\end{center} \caption{Average number of rounds of error correction before the
logical qubit is corrupted as a function of the physical gate error
rate $p_g$ and code distance $d$ when the separation $s$ of syndrome
changes is calculated taking the propagation of errors by two-qubit
gates into account.}\label{fig:dt_clean}
\end{figure}

The separation of any given pair of syndrome changes is now
calculated by determining the number of edges in the shortest path
connecting them. When all gates take the same time, the ideal
execution order remains as shown in Fig.~\ref{fig:figure2c}b.
Nevertheless, as can be seen in Fig.~\ref{fig:dt_clean}, the new
method of calculating the separation, a modification of the
classical processing only, dramatically changes the performance of
surface code QEC. The threshold error rate remains unchanged at
approximately 0.75\%. A lattice of distance $d$ can now reliably
correct $\lfloor (d-1)/2 \rfloor$ errors. For large $d$, the new
approach provides a quadratic improvement of the logical error rate.
Even for modest parameters, such as $p_g=10^{-4}$ and $d=7$, this
translates to a logical error rate improvement of over two orders of
magnitude.

\section{Conclusion}
\label{Conclusion}

By accurately tracking the propagation of errors due to the
two-qubit gates use during surface code QEC, we have improved the
number of errors a $d\times d$ code can reliably correct from
$\lfloor (d-1)/4 \rfloor$ to $\lfloor (d-1)/2 \rfloor$, which is
optimal. Explicitly, given a gate error rate $p_g=10^{-4}$ and a
surface code logical qubit with $d=7$ (a $13\times 13$ physical
qubit lattice), our method leads to over two orders of magnitude
improvement in the reliability of the logical qubit. This
improvement is purely as a result of better classical processing. In
the limit of large $d$, the new approach provides a quadratic
improvement of the logical error rate.

\section{Acknowledgements}

We acknowledge support from the Australian Research Council, the
Australian Government, and the US National Security Agency (NSA) and
the Army Research Office (ARO) under contract number
W911NF-08-1-0527.

\bibliography{../../../References}

\end{document}